\documentclass[aps,pre,preprint,groupedaddress,showpacs]{revtex4}
\usepackage{graphicx}
\usepackage{amssymb}
\usepackage{amsmath}

\begin{document}
\title{Three dimensional instability of flexible ferromagnetic filament loop}
\author{K.\={E}rglis}
\affiliation{University of Latvia, Ze\c{l}\c{l}u-8, R\={\i}ga, LV-1002, Latvia}
\author{R.Livanovi\v{c}s}
\affiliation{University of Latvia, Ze\c{l}\c{l}u-8, R\={\i}ga, LV-1002, Latvia}
\author{A.C\={e}bers}
\email[]{aceb@tesla.sal.lv}
\affiliation{Institute of Physics, University of Latvia, Salaspils-1, LV-2169, Latvia}

\date{\today}
\begin{abstract}
Dynamics of flexible ferromagnetic filaments in an external magnetic field is considered.
We report the existence of a buckling instability of the ferromagnetic filament at the magnetic field reversion, which leads to the formation of a metastable loop.
Its relaxation through three dimensional transformation of the configurations is observed experimentally and confirmed by numerical simulations. Bending modulus of the flexible ferromagnetic filaments synthesized by linking micron size core-shell ferromagnetic particles with DNA fragments is estimated by comparison of the parameters of the loops observed in the experiment with theoretical calculations. Formation of the loop and its relaxation are characterized by the numerically calculated writhe number. The relaxation time of the loop allows us to estimate the hydrodynamic drag of the filament.
\end{abstract}


\maketitle
\section{Introduction}
Ferromagnetic filaments are used by magnetotactic bacteria for the purpose of navigation in the magnetic field of the Earth \cite{1,2}. They may be created artificially by linking commercially available functionalized ferromagnetic microparticles with biotinized fragments of DNA \cite{3}. Different phenomena are known for ferromagnetic filaments. Due to their flexibility ferromagnetic filaments orientate perpendicularly to the AC magnetic field if its frequency is high enough \cite{3}. At magnetic field inversion they form loops \cite{4}, which are metastable due to their migration to one of the ends of the filament.

At the present moment the attention of researchers is focused on the creation of micro devices which mimic  self-propelling microorganisms \cite{5}. It has been demonstrated that the chains of superparamagnetic particles may be driven by an AC magnetic field \cite{6,7,8,9,10}. An essential feature of these developments is the symmetry breaking of the filament which causes its self-propulsion. The symmetry breaking is achieved by linking some cargo to one end of the filament \cite{6,8,9}, considering the defects in the chain of the magnetic particles \cite{7,8} and inducing spatial symmetry breaking by the buckling instabilities of the magnetic rod \cite{7,8}. It is important to note that anisotropy of the hydrodynamic drag is necessary for the self-propulsion of the magnetically actuated swimmer \cite{6,7,8,9,10}. Neglect of anisotropy of the hydrodynamic drag invalidates different schemes proposed for the magnetically actuated microswimmer (see for example the discussion in \cite{11}). Due to their property of loop formation ferromagnetic filaments present an opportunity to consider a new mechanism of the actuation of the magnetic microswimmer \cite{12}. By periodically inducing the buckling instability of the filament it is possible to create a microdevice which mimics the swimming of the biflagelate algae \textit{Chlamydomonos reinhardi} \cite{13} with the characteristic power and recovery strokes. Recently these algae have attained interest from the point of view of the study of synchronization phenomena \cite{14}, the transition to multicellularity \cite{15} and other interesting phenomena.

Here the numerical algorithm based on \cite{16} is developed for the numerical study of the stability of the loop in the 3D case. By numerical simulations it is shown that besides the instability due to the migration of loop there is a three dimensional instability. This conclusion is supported by experimental observation of the loop dynamics at magnetic field inversion.
The formation of the loop and its relaxation are characterized by the numerical calculation of the writhe number $Wr$.
It is found that the characteristic lifetime of the loop depends on the magnitude of the perturbation - the closer the loop is to the planar configuration the longer its lifetime.

\section{Model and algorithm of numerical simulation}
Here we consider the buckling instability of the ferromagnetic filament induced by the reversion of the magnetic field. The physics of the process is the following. By applying a constant magnetic field a filament with the permanent magnetic moment orients along the field as a compass needle. At the magnetic field reversion the filament becomes unstable and relaxes to the energetically favorable state with the magnetization along the field. There are two pathways to how this may happen. One possibility is that both ends of the filament rotate in the same direction, for example clockwise. This so called S-like mode corresponds to the reorientation of the solid compass needle along the direction of the reversed magnetic field, which due to the flexibility of the filament occurs through intermediate S-like shapes. This mode of the filament relaxation does not have a threshold value of the magnetoelastic number $Cm=MHL^{2}/A$, where $M$ is the magnetization per unit length of the filament, $2L$ is its length, and $A$ is the bending modulus. Besides this the filament at $Cm>\pi^{2}/4$ \cite{17} has the relaxation mode when the both free ends rotate in opposite directions as is illustrated in Fig.~1.
In the nonlinear stage development of this mode leads to the formation of the loop. It is metastable since as shown in the frame of the two-dimensional model \cite{4} it relaxes by migration to one of the free ends of the filament. The characteristic relaxation time grows exponentially with the magnetoelastic number $Cm$ and thus according to the two-dimensional model the loop is a long living metastable state.

We illustrate here that there is a more effective process of loop relaxation through the third dimension. Its physics is the following, - for the loop the magnetization of its free ends is along the magnetic field and the bent fragment of the filament has a magnetization component antiparallel to the field. The magnetization of this fragment relaxes to the direction of the magnetic field by bending out of the plane of the loop. Necessary symmetry breaking for out of plane deformations arises due to random perturbations.

Since by the formation of the loop the symmetry in the plane of the bending is broken then by inducing the periodic buckling of the filament in the AC magnetic field it is possible to create a self-propelling microdevice as it is illustrated in \cite{12}. For the self-propulsion of the ferromagnetic filament the process of the loop relaxation through the third dimension may impose definite conditions on the time dependence of the magnetic field since the symmetry of the filament will be restored by the relaxation through the third dimension and the self-propelling motion will stop. We believe that this may be overcome by choosing a corresponding protocol for the time dependence of the applied magnetic field.

The theoretical model for the dynamics of the ferromagnetic filament is developed in \cite{17,18,19} and is based on the expressions for the force $\vec {F}$ and torque $\vec {T}$ in the cross-section of the filament.
It is assumed that magnetization is along the tangent vector of the filament (see Fig.~1). If the directions of the tangent do not coincide the torques arise and the filament may buckle. In the Frenet frame the relations for $\vec{F}$ and $\vec{T}$ read
\begin{equation}
\vec{F}=A\frac{d}{dl}\Bigl(\frac{1}{R}\Bigr)\vec{n}+A\frac{1}{R}\tau\vec{b}+(\Lambda-\frac{1}{2}A\frac{1}{R^{2}}) \vec{t}+M(\vec{H}-\vec{t}(\vec{t}\cdot\vec{H}))~;
\label{Eq:1}
\end{equation}
\begin{equation}
\vec{T}=-A\frac{1}{R}\vec{b}~,
\label{Eq:2}
\end{equation}
where $\vec{t}$ is the tangent vector to the center line of the filament, $\vec{n},\vec{b}$ are the normal and binormal of the center line respectively, $l$ is the arclength of the center line, $R$ is the radius of the curvature of the center line and $\tau$ its torsion. $\Lambda$ characterizes tension arising due to the inextensibility of the filament, $-M\vec{t}$ is the magnetization per unit length of the filament, $\vec{H}$ is the magnetic field strength. The contribution of the twist is neglected in (\ref{Eq:1}) and (\ref{Eq:2}). Different effects of the twist on the ferromagnetic filaments are considered in \cite{19}.

In the simplest case the dynamics of the filament is described in the Rouse approximation
\begin{equation}
\zeta \vec{v}=\frac{d\vec{F}}{dl}~.
\label {Eq:3}
\end{equation}
In the equilibrium $\vec{F}=0$ if the external forces are absent. For the planar configuration introducing angle $\vartheta$, which the tangent makes with the field direction $\vec{t}=(\cos{(\vartheta)},\sin{(\vartheta)})$, the condition $\vec{F}\cdot \vec{n}=0$ leads to the equation
\begin{equation}
A\frac{d^{2}\vartheta}{dl^{2}}+MH\sin{(\vartheta)}=0~,
\label{Eq:4}
\end{equation}
which in the case of the unclamped ends at $l=\pm L$ is solved at boundary conditions $d\vartheta/dl\mid_{l=\pm L}=0$.
The solution reads
\begin{equation}
\vartheta=-2\arcsin{\Bigl(\sin{\Bigl(\frac{\vartheta_{m}}{2}\Bigr)}
sn\Bigl(\frac{\sqrt{Cm}l}{L},\sin^{2}{\Bigl(\frac{\vartheta_{m}}{2}\Bigr)}\Bigr)\Bigr)}~.
\label{Eq:5}
\end{equation}
Here $\pm \vartheta_{m}$ are the tangent angles at the free ends of the filament and $sn(x,k^{2})$ is the Jacobi elliptic function. The angle $\vartheta_{m}$ is found from the solution of the equation
\begin{equation}
K\Bigl(\sin^{2}{\Bigl(\frac{\vartheta_{m}}{2}\Bigr)}\Bigr)=\sqrt{Cm}~,
\label{Eq:6}
\end{equation}
here $K(k^{2})$ is elliptic integral of the first kind.
Solution (\ref{Eq:5}) for the maximal curvature of the filament at its center gives
\begin{equation}
\Bigl(\frac{L}{R}\Bigr)_{max}=2\sin{\Bigl(\frac{\vartheta_{m}}{2}\Bigr)}\sqrt{Cm}~.
\label{Eq:7}
\end{equation}
The dependence of the maximal curvature on $Cm$ determined by the relation (\ref{Eq:7}) is shown in Fig.~2. The Fig.~2 shows that at $Cm\geq 10$ the maximal curvature is well described by the asymptotic relation at $\vartheta_{m}\rightarrow \pi$ $(L/R)_{max}=2\sqrt{Cm}$. Dependence shown in Fig.~2 may be used to determine the elastic properties of the ferromagnetic filaments by measurements of their curvature. The elastic properties of superparamagnetic filaments linked by PAA in the similar way are determined in \cite{20}.

Using the Frenet equations $d\vec{t}/dl=-\vec{n}/R$ and $d\vec{n}/dl=\vec{t}/R+\tau \vec {b}$ the relation (\ref{Eq:1}) may be rewritten as follows
\begin{equation}
\vec{F}=-A\frac{d^{3}\vec{r}}{dl^{3}}+M\vec{H}+\tilde{\Lambda}\vec{t}~,
\end{equation}
where $\tilde{\Lambda}=\Lambda-M\vec{t}\cdot\vec{H}-3A/2R^{2}$. Tildes further are omitted.

The set of equations is solved numerically, discretizing the filament by a series of $n+1$ segments with the length $h$
($\alpha=(\zeta h)^{-1}$ is mobility of segment)
\begin{equation}
\vec{v_{i}}=\alpha(\vec{f_{i}}+\vec{f'_{i}}); (i=0,...,n+1)~,
\end{equation}
here
\begin{eqnarray}
\vec{f_{i}}=-hAd^{4}\vec{r}/dl^{4}\mid_{i};\quad \vec{f'_{i}}=hd(\Lambda\vec{t})/dl\mid_{i}\quad (i=1,...n-1)~; \\ \nonumber
\vec{f_{0}}=-Ad^{3}\vec{r}/dl^{3}\mid_{0}+M\vec{H};\quad \vec{f'}_{0}=\Lambda\vec{t}\mid_{0};\\ \nonumber \vec{f_{n}}=Ad^{3}\vec{r}/dl^{3}\mid_{n}-M\vec{H};\quad \vec{f'}_{n}=-\Lambda\vec{t}\mid_{n}
\end{eqnarray}
are the forces on the segments due to their bending, magnetic torque and tension.

The stress $\Lambda \vec{t}$ is found using the condition of the inextensibility $\vec{t}d\vec{v}/dl=0$. It in the integral form gives
\begin{equation}
0=-\int \Lambda\vec{t}\frac{d\vec{v}}{dl}dl=\Lambda\vec{t}\mid_{0}\vec{v_{0}}-\Lambda\vec{t}\mid_{n}\vec{v_{n}}
+\int\vec{v}\frac{d(\Lambda\vec{t})}{dl}dl~.
\label{Eq:inc}
\end{equation}
In the discrete form the condition (\ref{Eq:inc}) reads
\begin{equation}
\vec{f'_{0}}\vec{v_{0}}+\vec{f'_{n}}\vec{v_{n}}+\sum_{i=1}^{n-1}f'_{i}\vec{v_{i}}=0~.
\label{Eq:8}
\end{equation}
The condition of inextensibility is imposed by $n$ constraints
$g_{k}=(\vec{r}_{k+1}-\vec{r}_{k})^{2}=h^{2}$ ($k=0,...,n-1;j=0,...n$).
The $n\times d(n+1)$ Jacobian matrix with elements $J=\partial g_{k}/\partial \vec{r_{j}}$
is introduced.
The relation (\ref{Eq:8}) put in the matrix form as $\vec{f'}\cdot\vec{v}=0$ is satisfied for allowed motions
$J\cdot\vec{v}=0$ by $\vec{f'}=J^{T}\gamma$ (upper script $^{T}$ denotes transposed matrix, $\gamma$ is $n$-dimensional colon vector) \cite{16}.
Equation of motion $\vec{v}=\alpha(\vec{f}+\vec{f'})$ accounting for constraints gives $J\vec{f}+J\vec{f}'=\alpha^{-1}J\vec{v}=0$ and thus $\gamma=-(J\cdot J^{T})^{-1}J\vec{f}$. As  a result the equation of motion reads
\begin{equation}
\vec{v}=\alpha P\vec{f}~,
\end{equation}
where $P=I-J^{T}(J\cdot J^{T})^{-1}J$ is the projection operator on the space of allowed motions ($P^{2}=P$).

Approximating $d^{4}\vec{r}/dl^{4}\mid_{i};(i=1,...,n-1)$ by the central differences, $d^{3}\vec{r}/dl^{3}\mid_{0}=(\vec{r}_{2}+\vec{r}_{0}-2\vec{r}_{1})/h^{3}$ and
$d^{3}\vec{r}/dl^{3}\mid_{n}=-(\vec{r}_{n-2}+\vec{r}_{n}-2\vec{r}_{n-1})/h^{3}$ the stiffness matrix $C$, which determines the elastic stresses ($\vec{f}=C\vec{r}+\vec f^{m}$) is constructed. The implicit scheme for the time step $\tau$ then gives
\begin{equation}
(I-\alpha\tau P\cdot C)(\vec{r}_{t+\tau}-\vec{r}_{t})=\alpha\tau P\cdot(C\vec{r}_{t}+\vec{f}^{m})~,
\end{equation}
here $\vec{f}^{m}_{0}=M\vec{H};\vec{f}^{m}_{n}=-M\vec{H};\vec{f}^{m}_{i}=0\quad (i=1,...,n-1)$.
The equation for $\vec{r}_{t+\tau}-\vec{r}_{t}$ is easily resolved by inverting the matrix $I-\alpha\tau P\cdot C$.

At the end of each time step the filament is reshaped to satisfy the constraints according to the algorithm described in \cite{16}. To put the equations in dimensionless form the following scales are introduced: time - $\zeta(2L)^{4}/A$ - the characteristic elastic relaxation time, length - 2L, and the elastic force $A/(2L)^{3}$. The only parameter which controls the dynamics of the filament in this scaling is the magnetoelastic number $Cm=MHL^{2}/A$.
\section{Dynamics of filaments}
The synthesis of the ferromagnetic filaments by linking functionalized with streptavidin core-shell ferromagnetic particles with size $d=4.26 \quad \mu m$ (Spherotech) with 1 kb long DNA fragments is carried out as described before \cite{3}. Magnetic properties of the diluted ensemble of these particles are measured by the vibrating sample magnetometer and are shown in Fig.~3. For the range of the magnetic field $H<200 \quad Oe$ used in our experiments the magnetization curve in Fig.~3 gives the value of the magnetic moment of the particle $m= 1.4\cdot 10^{-10}\quad emu $. The linear density of the magnetization $M=m/d$ then is $M=3.3\cdot 10^{-7} \quad emu$.

The protocol of the experiment is the following. The magnetic field is created by two coils in the space between two microscope slides at a distance of $50~\mu m$. The coils are supplied by power supply-amplifier Kepco BOP 20-10M. A constant magnetic field is applied and the filament is oriented along the field lines. Then by switching the direction of the current in the coils in time less than $4~ms$ determined by the inductance of the coils the magnetic field is reversed. The dynamics of the configurations of the filament is registered by the video camera Mikrotron MC1363 at 50 frames per second.

A sequence of several configurations of the ferromagnetic filament under the field inversion is shown in Fig.~4. It illustrates how the loop formed at an intermediate stage relaxes through transformation to a three dimensional configuration. The observed dynamics is compared with results of numerical simulations. Coordinates of the filament projection on $x,y$ plane are obtained from the sequence of video frames by ImageJ. The contour length of the projection is calculated and the coordinates of the filament are smoothed by the polynomial of degree 11. Dimensionless coordinates of the filament are obtained by normalizing with the contour length of the initial configuration. The comparison of the results of the numerical calculations at with the experimental observations is shown in Fig.~5.
To obtain the quantitative agreement of the parameters of the loop observed in experiment with numerical simulation it is carried out at $Cm=5.5$.  As initial configuration for the numerical simulation the initial configuration of the filament for the symmetry breaking inclined at angle $\alpha$ ($\tan{\alpha}=10^{-4}$) to $x,y$ plane is taken. Numerical simulation in agreement with the experiment shows that the filament makes a loop and does not relax to the straight configuration along the reversed direction of the field by the S-like deformation mode, which might develop at corresponding initial conditions \cite{17}.
The qualitative and for last 3-4 configurations quantitative agreement of the configurations obtained in experiment and numerical simulation is observed. The qualitative differences with the experiment start from the last configuration. In the numerical experiment filament through the third dimension rather fast relaxes, as it is illustrated in Fig.~6, to the straight configuration along the direction of magnetic field. In the experiment the delay of the loop relaxation is observed and the end of the loop slides along its other leg for the time period 0.3 s, twice as large as the time necessary of the loop formation. We presume that it is due to the magnetic interaction of the end of the loop with the filament which is not taken into account in our model. The transition to the delayed relaxation stage of the loop is illustrated by the time dependence of the curvature of the filament in Fig.~7. The curvature is calculated by differentiation of the smoothing polynomial of degree 11. The error bars are estimated from 5-10 percent error in the coordinates of the filament due to their smoothing by the polynomial. To take into account the third dimension the curvature of the projection is corrected by the multiplier $L_{p}/L_{0}$, where $L_{p}$ is the loop projection length and $L_{0}$ is the length of the filament. Transition to the delayed relaxation stage takes place at 0.2 s and corresponds to the last configuration shown in Fig.~5. Relation (7) at $Cm=5.5$ for the parameter $(L/R)_{max}$ gives the value 4.3 not far from the value 4.85 corresponding to the final stage of loop formation shown in Fig.~5.
For comparison with the theoretical curve of the curvature of the filament in its center given in a two dimensional model by the relation (\ref{Eq:7}) in Fig.~8 we give the maximal curvature of the three dimensional filament and its curvature in the center obtained by numerical calculations at $\tan{\alpha}=10^{-3}$. Corresponding curvature is given for the time moment when it is the maximal in a time sequence of the configurations for the given magnetoelastic number. We see that the curvature of the two dimensional model is between the maximal curvature and the curvature in the center of the filament.

The value of the magnetoelastic number since the the linear magnetization density is known allows us to estimate the bending modulus of the filament $A=1.5\cdot 10^{-10}\quad erg\cdot cm$. The characteristic time of the loop formation shown in Fig.~4 is $0.2~s$ and allows us to estimate the hydrodynamic drag coefficient $\zeta=4\pi\eta=0.54\quad g\cdot cm^{-1}\cdot s^{-1}$ which corresponds to the viscosity $0.043\quad g\cdot cm^{-1}\cdot s^{-1}$ approximately 4 times larger as the viscosity of water. This indicates that the hydrodynamic interaction of the filament with walls of the layer not taken into account in our model is important. Another indication about this are given by numerical simulations taking into account the anisotropy of hydrodynamic drag. It is incorporated introducing in the left side of equation (3) the matrix $B^{-1}=I+(\zeta_{\parallel}-\zeta_{\perp})\zeta_{\perp}^{-1}\vec{t}\otimes\vec{t}$. Numerical calculation shows that at $(\zeta_{\parallel}-\zeta_{\perp})\zeta_{\perp}^{-1}=-1$, which corresponds to the filament in unconfined conditions, relaxation to the straight configuration is more fast as it should be for the given value of the magnetoelastic number.
The dynamics at $(\zeta_{\parallel}-\zeta_{\perp})\zeta_{\perp}^{-1}=1$ is close to shown in Fig.~5.

Besides the issues of hydrodynamic and magnetic interactions which have not taken into account by our model we should mention that the filament may possess different defects which may influence its dynamics \cite{8}. This may explain the quantitative differences of the filament evolution with the numerical simulation which may be seen in Fig.~5.

The dynamics during the loop formation and its relaxation may be characterized by the writhe $Wr$ \cite{21}, which may be calculated according to the Fuller formula \cite{22}
\begin{equation}
Wr=\frac{1}{2\pi}\int \frac{\vec{e}_{z}(d\vec{t}\times \vec{t}/dl)}{1+\cos{(\vartheta)}}dl~,
\end{equation}
where $\vartheta$ is the angle which the tangent to the filament makes with the $z$ axis.
Time dependence of $Wr$ for different inclination angles $\alpha$ of the plane of initial bending to the field at $Cm=6$ is shown in Fig.~9. We see that the smaller the inclination angle the longer the loop lifetime. The sign of the $Wr$ is determined by the sign of the inclination angle.
Inclining the plane of initial bending in opposite direction $Wr$ is positive.

The loop formation and its relaxation at larger values of the magnetoelastic number has quantitative differences. As it is illustrated in Fig.~10 for $Cm=25$ the loop formed at an intermediate stage has smaller radius, what is expected according to relation (\ref{Eq:7}).
We draw attention that the filament in Fig.~6 similarly to what is observed in the experiment (Fig.~3(3)) has configuration close to the ring what is not the case at $Cm=25$ shown in Fig.~10, where the free ends are much longer.
This confirms our estimate of the magnetoelastic number for the loop observed in the experiment.

\section{Conclusions and discussion}
To conclude it is shown that the loops of ferromagnetic filaments formed at magnetic field inversion are metastable and relax through the transformation to three dimensional configurations. Dynamics of the loop formation and relaxation is characterized by writhe calculations which show that the lifetime of the loop depends on the value of the initial perturbation. Bending modulus of ferromagnetic filaments synthesized by linking the core-shell micron size ferromagnetic particles with the DNA fragments is estimated by the comparison of the characteristics of the loops observed in the experiment with the theoretical calculations.

\begin{acknowledgements}
Authors are thankful to M.M.Maiorov for his assistance in taking the magnetization curve of the suspension of ferromagnetic particles. The work is supported by the grant of University of Latvia ESS2009/86.
\end{acknowledgements}

\begin{figure}
 \includegraphics [width=0.5\textwidth] {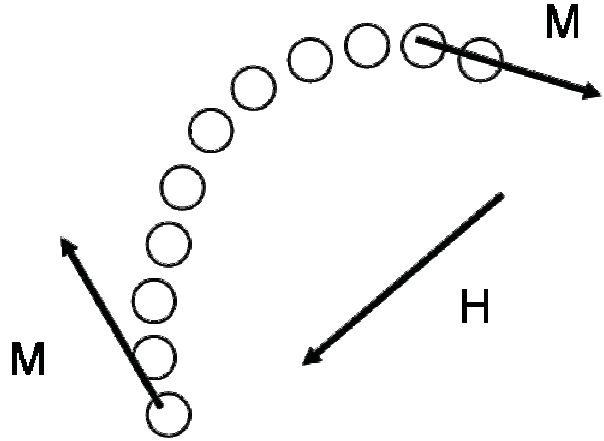}
 \caption{Cartoon of a ferromagnetic filament at its bending in an external field.}
 \end{figure}

\begin{figure}
 \includegraphics [angle=-90,width=1.0\textwidth] {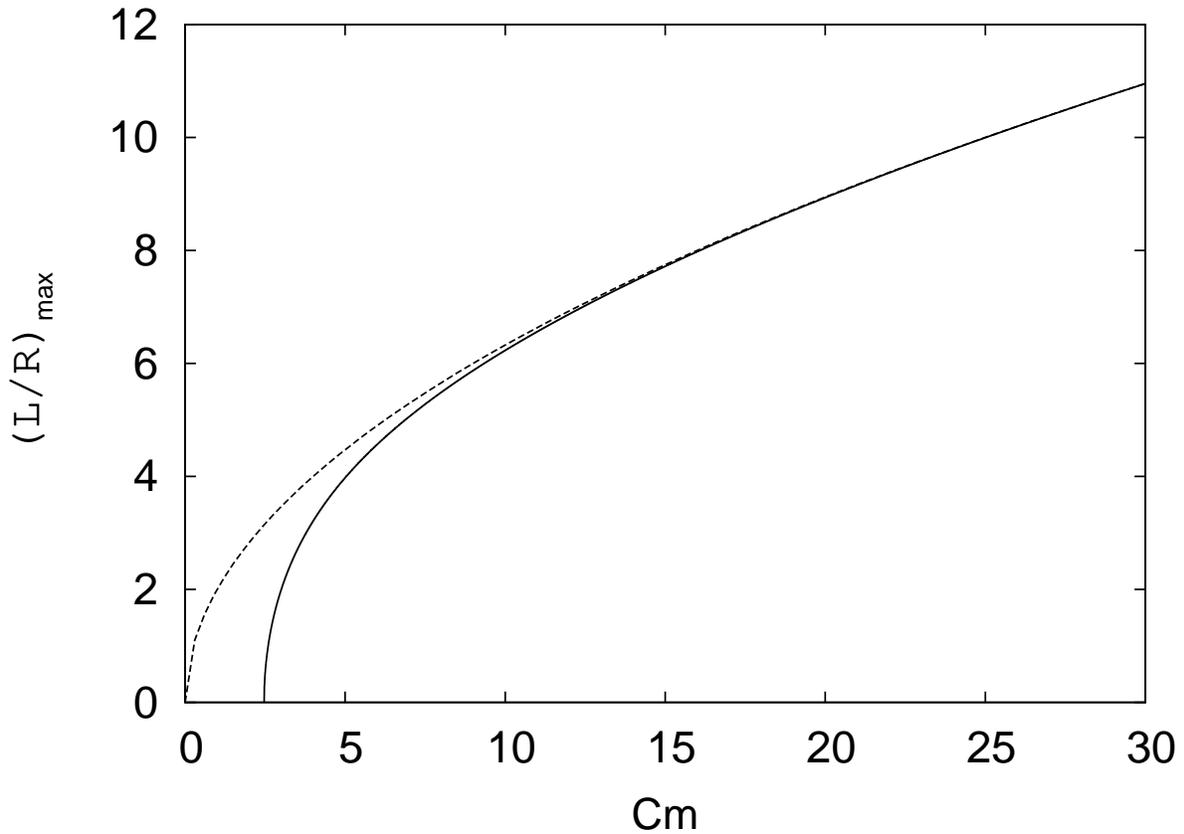}
 \caption{Curvature of ferromagnetic filament in dependence on the magnetoelastic number. Solid line - theoretical dependence (\ref{Eq:7}), dashed line - asymptotic relation $(L/R)_{max}=2\sqrt{Cm}$. Solid line intersects x axis at the critical value of the magnetoelastic number for the bending instability of the filament $Cm_{c}=\pi^{2}/4$.}
 \end{figure}

 \begin{figure}
 \includegraphics [angle=-90,width=1.0\textwidth] {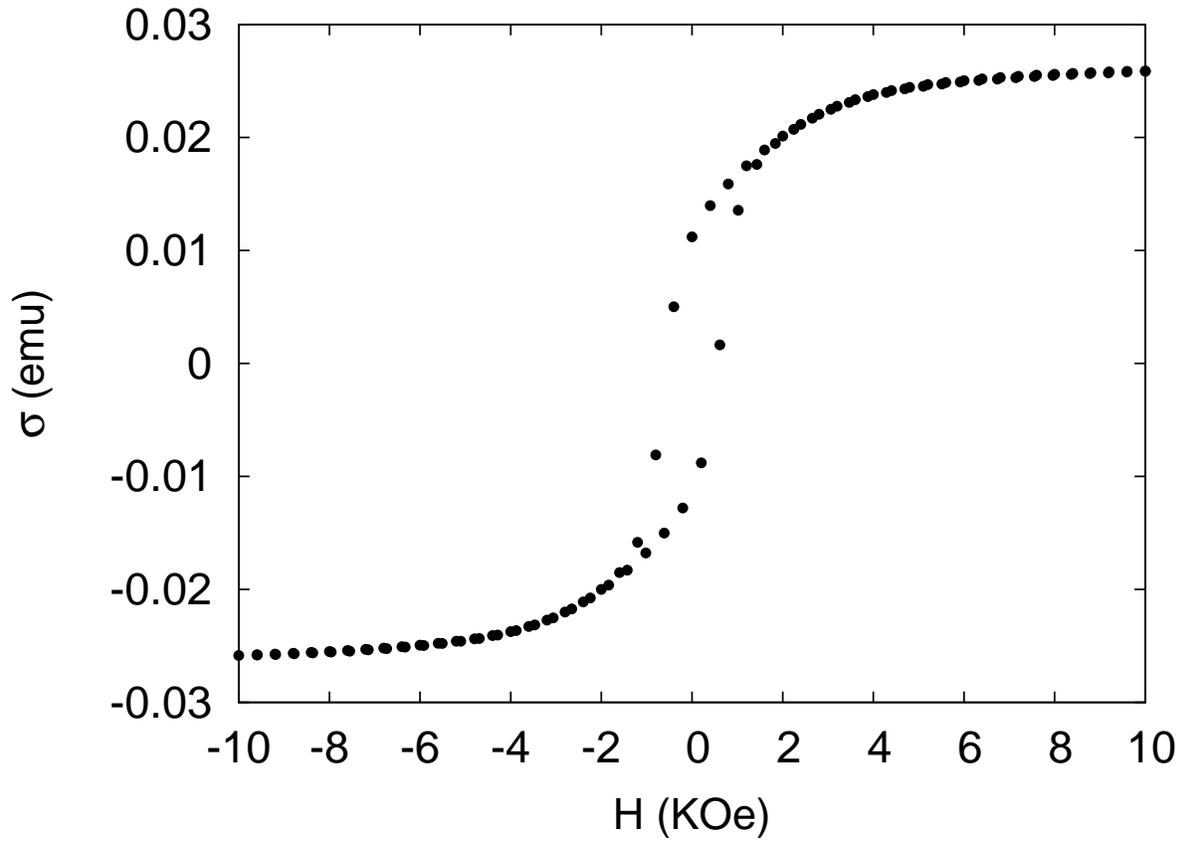}
 \caption{Magnetic moment of the diluted sample of core-shell ferromagnetic particles in dependence on the magnetic field. Mass of sample $m=3.64\quad mg$.}
 \end{figure}

 \begin{figure}
 \includegraphics [width=1.0\textwidth] {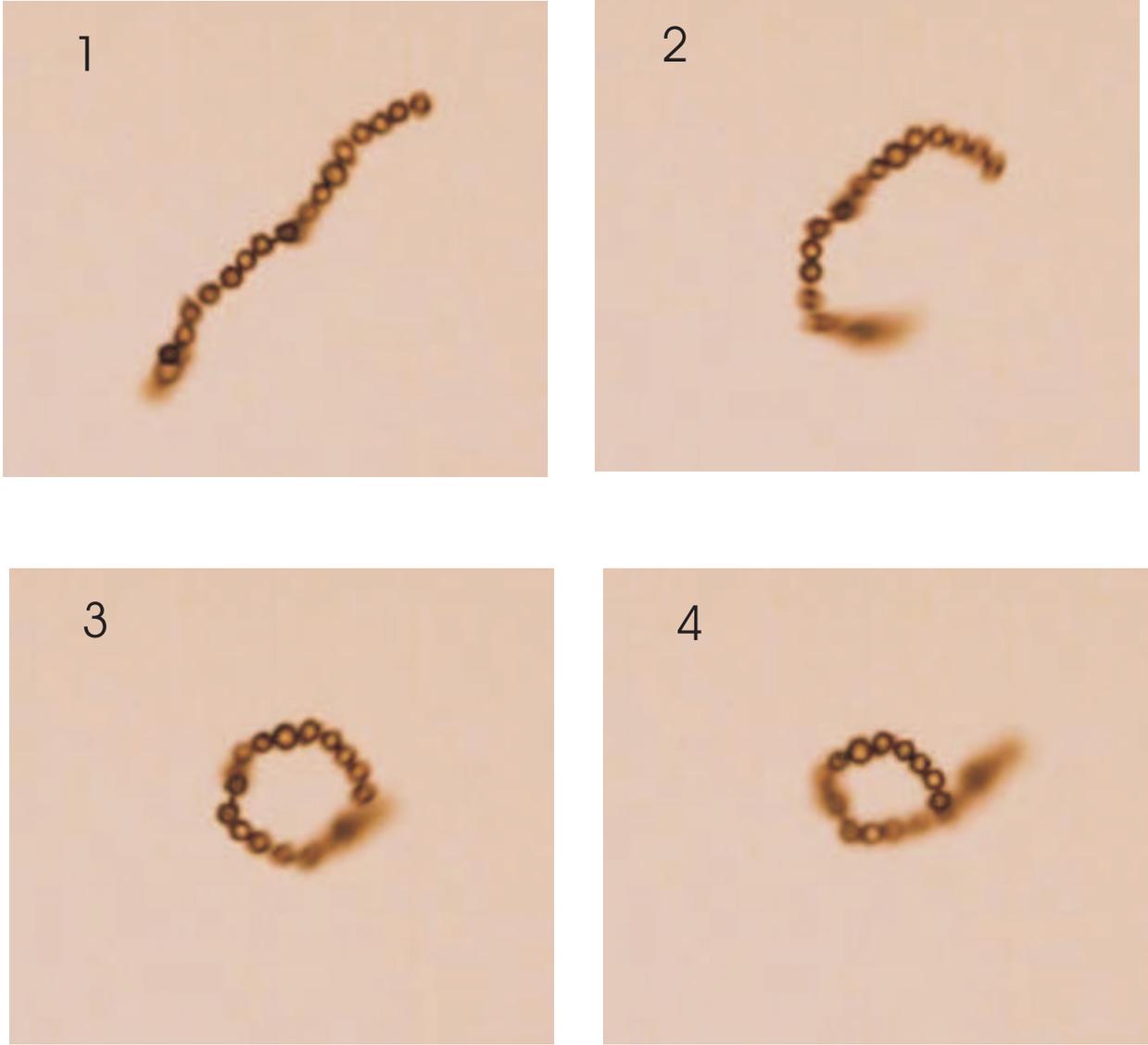}
 \caption{Loop formation and relaxation due to the escape into the third dimension at the inversion of the magnetic field. Time - 0(1);0.1 s(2); 0.16 s(3);0.28 s (4). $H=180\quad Oe$,the length of the filament $2L=74.6\quad \mu m$.
 Field makes angle $\pi/4$ with the horizontal direction.}
 \end{figure}

\begin{figure}
 \includegraphics [angle=-90,width=1.0\textwidth] {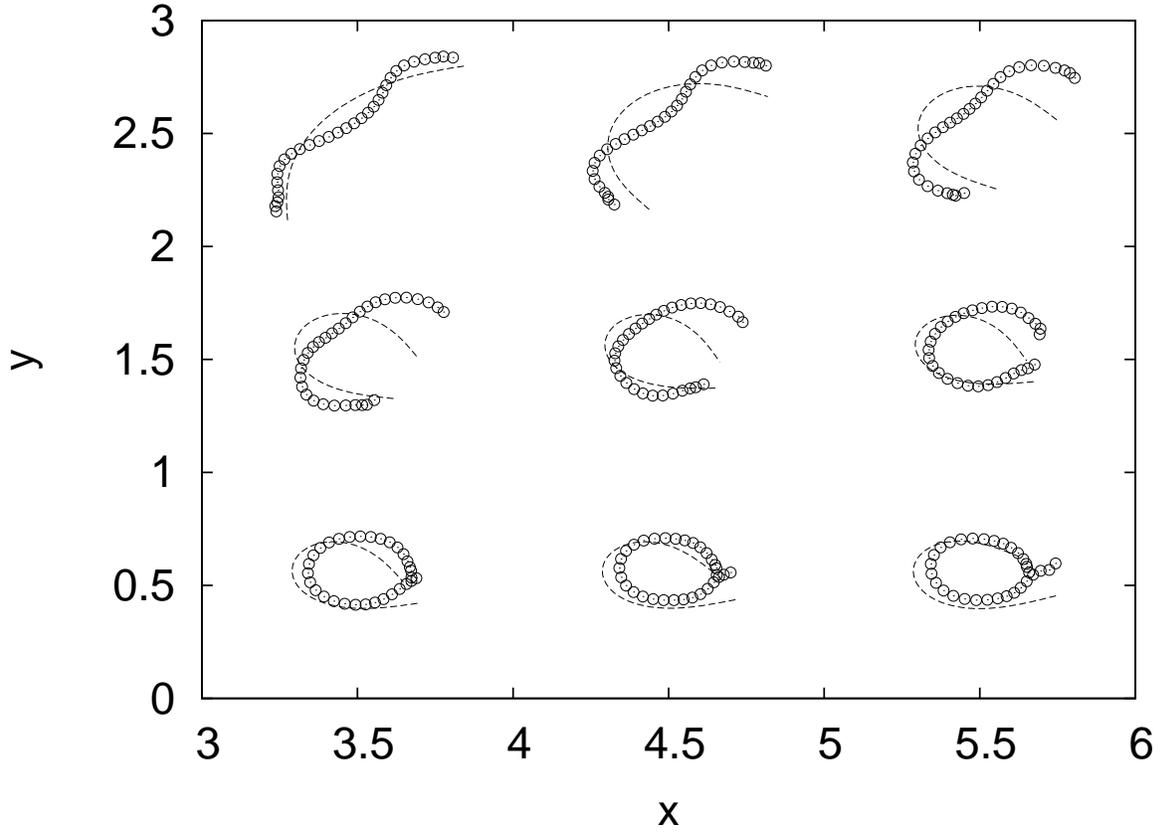}
 \caption{Comparison of experimentally observed loop formation with numerical simulation (dashed lines). Dimensionless time starting from upper left corner increases with step 0.002 starting from 0.002. $Cm=5.5$. Observation time - increases by 0.02 s starting from 0.02 s. $H=180\quad Oe$,the length of the filament $2L=74.6\quad \mu m$.
 Field makes angle $\pi/4$ with the x axis direction.}
 \end{figure}

\begin{figure}
 \includegraphics [width=1.0\textwidth] {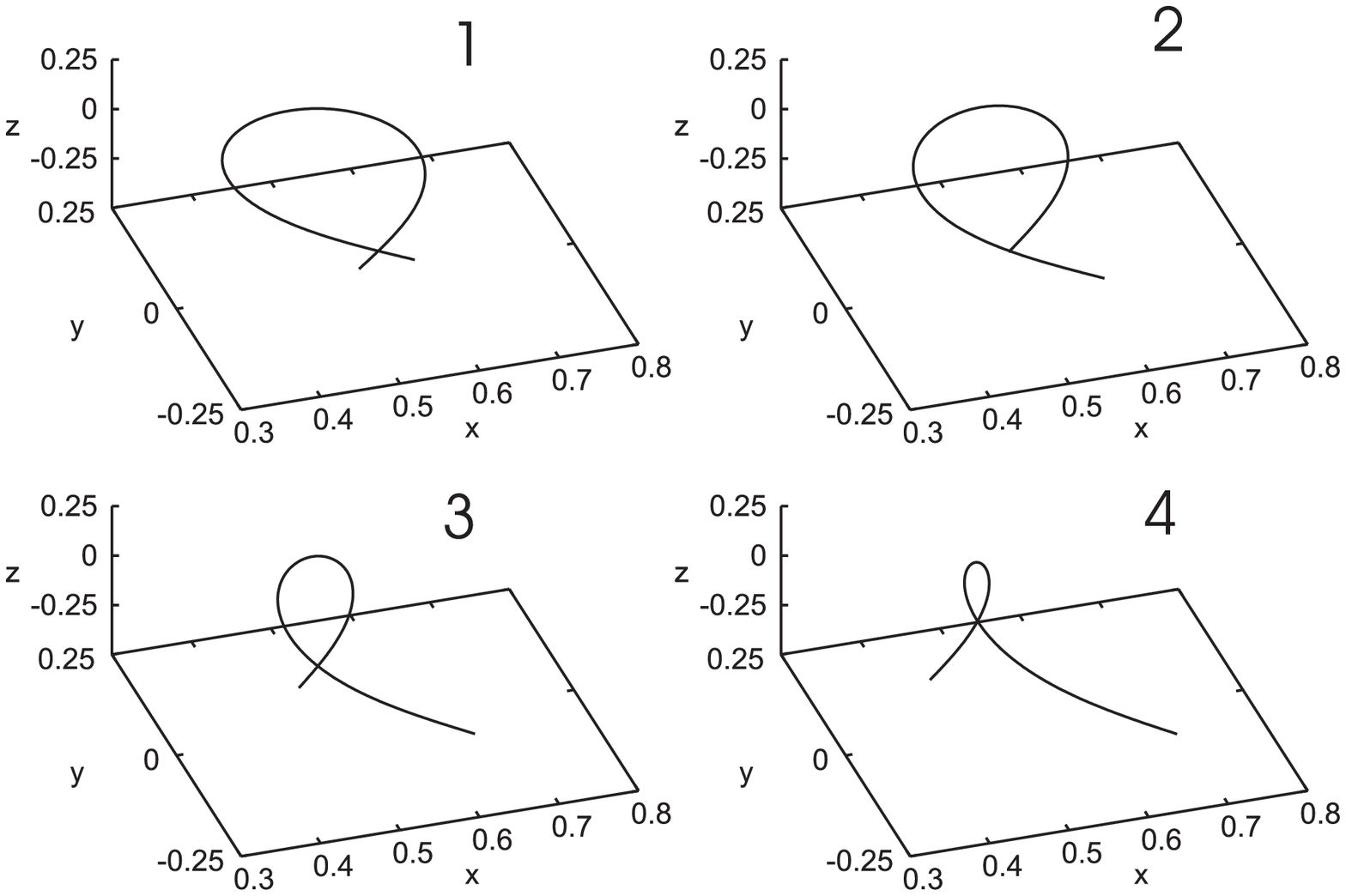}
 \caption{Dynamics of the loop relaxation as given by numerical simulation. Time in dimensionless units - 0.02(1);0.029(2);0.03(3);0.0305(4). $Cm=6,\tan{(\alpha)}=10^{-7}$. Initial configuration $y=\varepsilon\sin{(\pi x/l)}(l=1-(\pi\varepsilon/2)^{2};\varepsilon=10^{-2})$.}
 \end{figure}

\begin{figure}
 \includegraphics [angle=-90,width=1.0\textwidth] {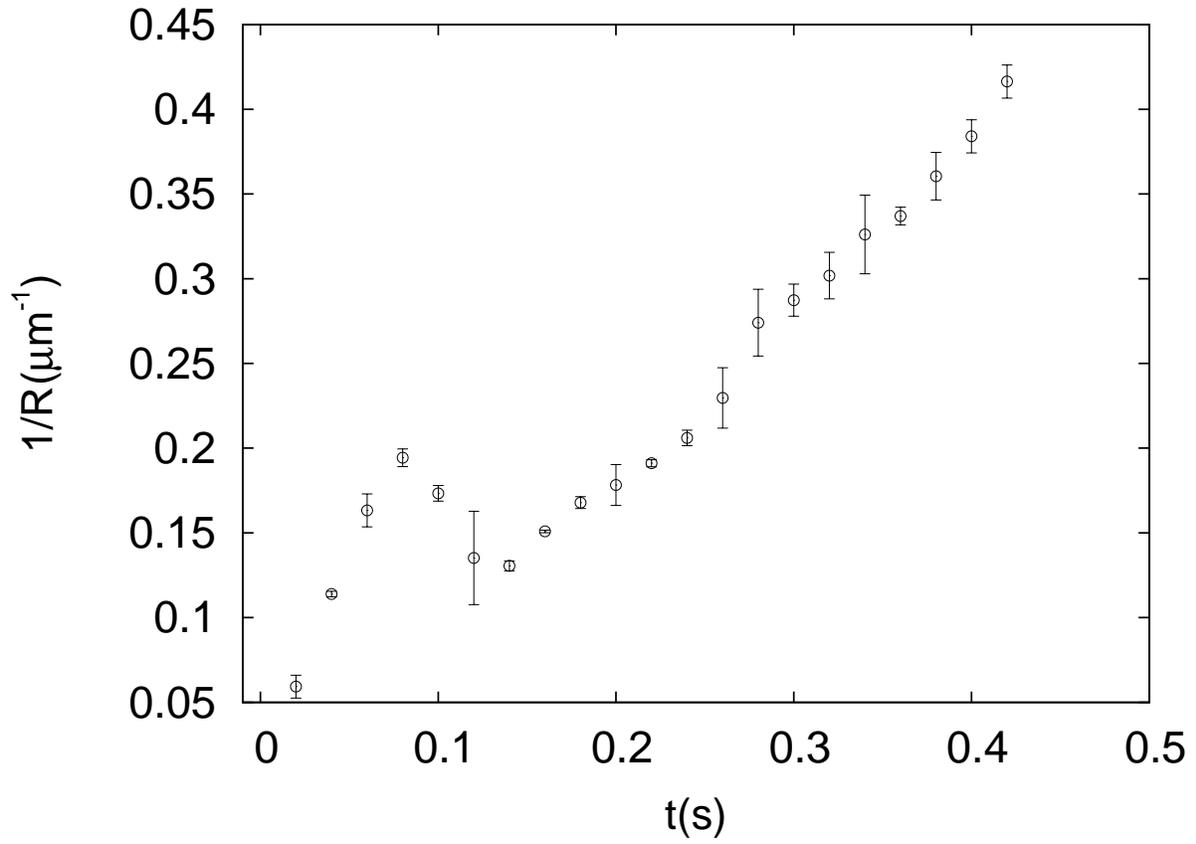}
 \caption{Curvature of the projection of the filament at the loop formation and delayed stage of its relaxation before the escape into the third dimension. $H=180\quad Oe$, the length of the filament $2L=88.5\quad \mu m$.}
 \end{figure}

\begin{figure}
 \includegraphics [angle=-90,width=1.0\textwidth] {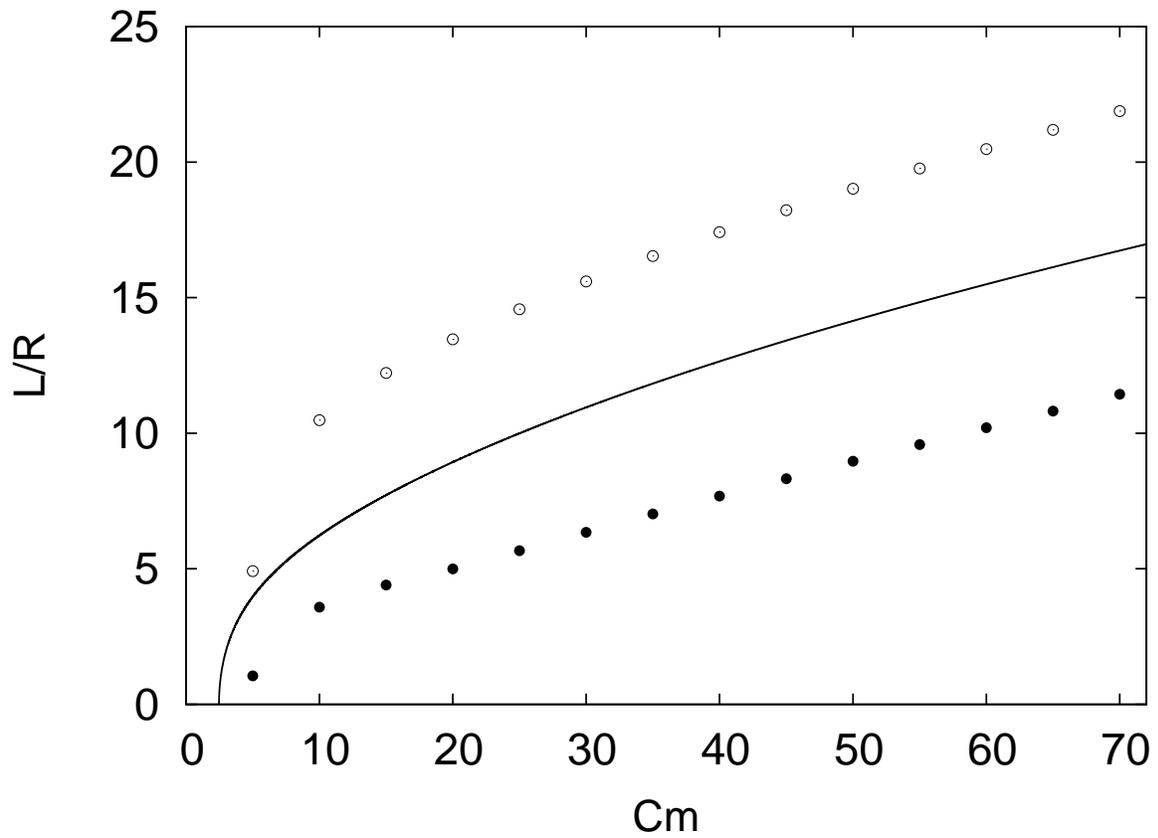}
 \caption{Maximal curvature (empty circles), curvature in the center (filled circles) in dependence on the magnetoelastic number $Cm$.
 $\tan{(\alpha)}=10^{-3}$. Solid line is the theoretical dependence (\ref{Eq:7}).}
 \end{figure}

 \begin{figure}
 \includegraphics [angle=-90,width=1.0\textwidth] {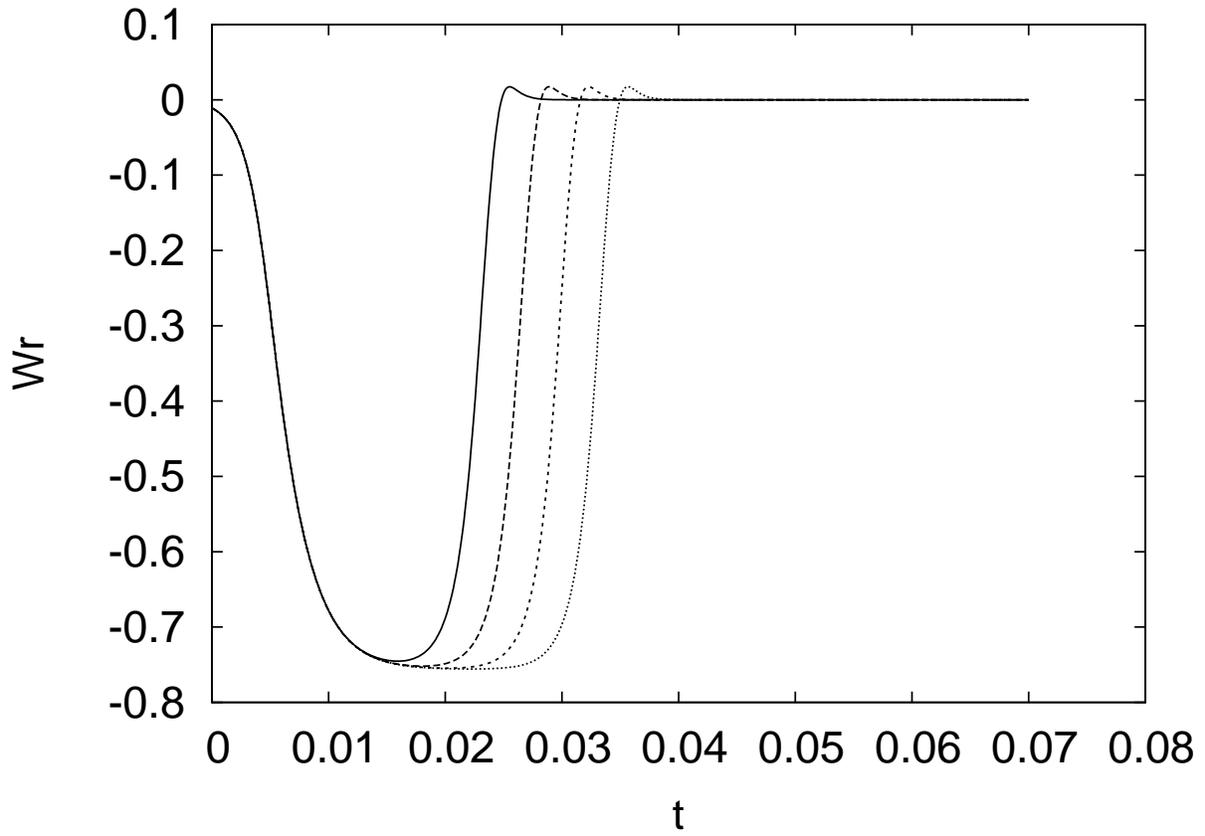}
 \caption{Time dependence of the $Wr$ for several values of the inclination angle $\alpha$ of the plane of initial bending ($\tan{(\alpha)}=10^{-5}$ (solid line);$10^{-6}$ (long dashed line);$10^{-7}$ (short dashed line);$10^{-8}$ (dotted line)). Lifetime of loop with smaller $\alpha$ is longer. $Cm=6$.}
 \end{figure}

 \begin{figure}
 \includegraphics [width=1.0\textwidth] {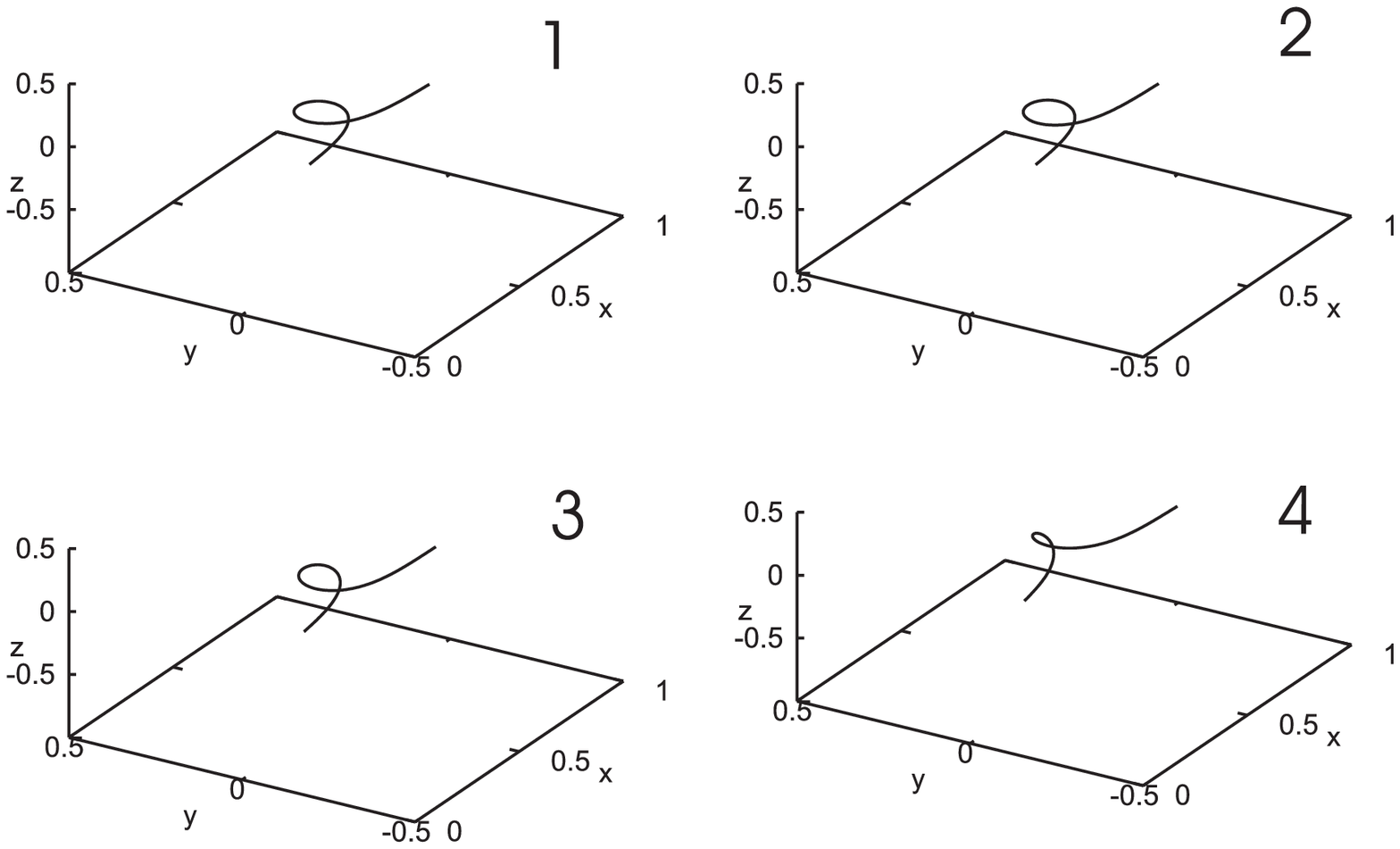}
 \caption{Dynamics of the loop relaxation as given by numerical simulation. Time in dimensionless units - 0.0034(1);0.0036(2);0.0038(3);0.0040(4). $Cm=25,\tan({\alpha})=10^{-4}$. Initial configuration $y=\varepsilon\sin{(\pi x/l)}(l=1-(\pi\varepsilon/2)^{2};\varepsilon=10^{-2})$.}
 \end{figure}

\end{document}